\documentclass[aps,prl,twocolumn,amsfonts,amssymb,amsmath,floats,floatfix,showpacs,preprintnumbers,superscriptaddress]{revtex4-1}

\usepackage{graphicx}
\usepackage{color}
\usepackage{dcolumn}
\usepackage{bm}
\usepackage{hyperref}
\usepackage{relsize}
\usepackage{soul}

\hypersetup{
colorlinks=true,
urlcolor= blue,
citecolor=blue,
linkcolor= blue,
bookmarks=true,
bookmarksopen=false,
}

\newcommand {\BiSe}{Bi$_2$Se$_3$}
\newcommand {\BiTe}{Bi$_2$Te$_3$}
\newcommand {\FeBiSe}{Fe$_x$Bi$_{2-x}$Se$_3$}
\newcommand {\FeBiTe}{Fe$_x$Bi$_{2-x}$Te$_3$}
\newcommand{\GM}{$\Gamma$-$M$}
\newcommand{\GK}{$\Gamma$-$K$}
\renewcommand{\vec}[1]{\bm{#1}}

\begin{document}

\title{Spin-Polarized Quantum Well States on Bi$_{2-x}$Fe$_x$Se$_3$}

\author{Michael M. Yee}
\thanks{The first two authors contributed equally to this work.}
\author{Z.-H. Zhu}
\thanks{The first two authors contributed equally to this work.}
\author{Anjan Soumyanarayanan}
\author{Yang He}
\author{Can-Li\ Song}
\affiliation{Department of Physics, Harvard University, Cambridge, MA 02138, USA}
\author{Ekaterina Pomjakushina}
\affiliation{Laboratory for Development and Methods, Paul Scherrer Institute, CH-5232 Villigen PSI, Switzerland}
\author{Zaher Salman}
\affiliation{Laboratory for Muon Spin Spectroscopy, Paul Scherrer Institute, CH-5232 Villigen PSI, Switzerland}
\author{Amit Kanigel}
\affiliation{Department of Physics, Technion-Israel Institute of Technology, Haifa 32000, Israel}
\author{Kouji Segawa}
\author{Yoichi Ando}
\affiliation{Institute of Scientific and Industrial Research, Osaka University, Ibaraki, Osaka 567-0047, Japan}
\author{Jennifer E. Hoffman}
\email{jhoffman@physics.harvard.edu}
\affiliation{Department of Physics, Harvard University, Cambridge, MA 02138, USA}

\date{\today}

\begin{abstract}
Low temperature scanning tunneling microscopy is used to image the doped topological insulator \FeBiSe. Interstitial Fe defects allow the detection of quasiparticle interference (QPI), and the reconstruction of the empty state band structure. Quantitative comparison between measured data and density functional theory calculations reveals the unexpected coexistence of quantum well states (QWS) with topological surface states (TSS) on the atomically clean surface of \FeBiSe. Spectroscopic measurements quantify the breakdown of linear dispersion due to hexagonal warping. Nonetheless, both QWS and TSS remain spin-polarized and protected from backscattering to almost 1\,eV above the Dirac point, suggesting their utility for spin-based applications.
\end{abstract}


\maketitle

Topological insulators (TIs), recently discovered materials with insulating bulk and topologically protected helical Dirac surface states, have generated widespread excitement due to proposed applications such as dissipationless spintronics, ambipolar transistors, and fault-tolerant quantum computers \cite{HasanRMP2010, QiRMP2011, AndoJPSJ2013}. Many of these applications hinge on the predicted ability to preserve spin information without backscattering. \BiSe\ has attracted particular attention due to the accessibility of its Dirac point within a relatively large $\sim$300 meV bulk band gap \cite{HsiehNature2009}, and the availability of additional quantum well surface states \cite{BianchiNatComm2010, WrayNatPhys2010, BeniaPRL2011, ZhuPRL2011, BianchiPRL2011, YeArxiv2011, ChenPNAS2012, VallaPRL2012, RoyPRL2014} which may be spin-polarized by a large Rashba effect \cite{KingPRL2011, BahramyNatCom2012}. However, spin-polarized quantum well states (QWS) depend sensitively on adsorbents, and have not been observed on clean surfaces. More generally, little is known about the empty state band structure of \BiSe, which is inaccessible to angle-resolved photoemission spectroscopy (ARPES). For both topological surface states (TSS) and QWS, it remains crucial to characterize the high energy extent to which they remain linearly dispersing, spin-polarized, and protected against surface-bulk scattering \cite{ParkPRB2010, KimPRL2011} and backscattering \cite{LeePRB2009}.
	
Scanning tunneling microscopy (STM) can provide real space images of both filled and empty states and their local relationship to surface and near-surface impurities. STM also provides access to momentum space information via quasiparticle interference (QPI) imaging. When quasiparticle states of energy $\varepsilon$ scatter elastically from impurities, the interference between initial and final quasiparticle wavevectors $\vec{k_i}$ and $\vec{k_f}$ can result in a standing wave pattern with wavevector $\vec{q} = \vec{k_f} - \vec{k_i}$ at energy $\varepsilon$. The observed dispersion of $\vec{q}(\varepsilon)$, and the inversion of $\vec{q}(\varepsilon)$ to find $\vec{k}(\varepsilon)$, has established QPI as a reliable $k$-space probe \cite{CrommieNature1993}. Indeed, QPI imaging has been used to demonstrate the protection against backscattering \cite{ZhangPRL2009, AlpichshevPRL2010} and the high energy breakdown of linear dispersion in pristine \BiTe\ \cite{SessiPRB2013}, as well as the onset of backscattering in \FeBiTe\ \cite{OkadaPRL2011}. However, dispersing QPI in \BiSe\ has been elusive \cite{HanaguriPRB2010, KimPRL2011, AlpichshevPRL2012}, appearing only in a limited energy range with inconsistent velocity \cite{WangPRB2011, BeidenkopfNatPhys2011, ZhangPRB2013}. Therefore, the high energy band structure and scattering mechanisms of \BiSe\ are unknown.

\begin{figure*} [t]
 \includegraphics[width=2\columnwidth,clip]{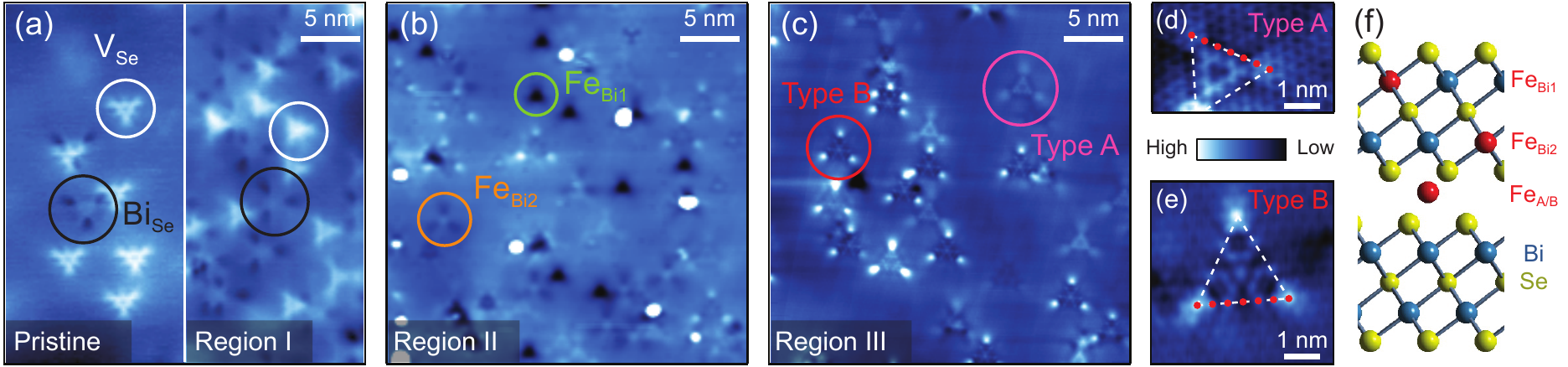}
  \caption{\label{fig:impurities} (color online). (a-c) STM Topographies of pristine \BiSe\ (a, left) and three regions of Fe$_{0.07}$Bi$_{1.93}$Se$_3$ [(a, right), (b), and (c)]. Six types of impurities, circled in (a)-(c), are segregated into three distinct micron-scale regions in Fe-doped samples. Region I is characterized by the same impurities seen in pristine samples, $i.e.$ Se vacancies (V$_{\mathrm{Se}}$) and Bi$_{\mathrm{Se}}$ antisites. Region II contains Fe substitutions at two different Bi sites (Fe$_{\mathrm{Bi1}}$ and Fe$_{\mathrm{Bi2}}$). Region III contains two types of Fe interstitials (type-A and type-B). (d), (e) High resolution topographies of type-A and type-B Fe interstitials. Their spatial signature has a similar vertex-to-vertex distance of 6$a_{0}$, indicated by red dots, where $a_{0}$ is the Se-Se lattice constant. (f) Crystal structure of \BiSe\ with Fe dopants shown in red. Experimental conditions: (a, left) $V_s$\,=\,0.4\,V, $I_s$\,=\,30\,pA; (a, right) $V_s$\,=\,0.5\,V, $I_s$\,=\,50\,pA; (b) $V_s$\,=\,1\,V, $I_s$\,=\,20\,pA; (c) $V_s$\,=\,0.55\,V, $I_s$\,=\,100\,pA; (d, e) $V_s$\,=\,0.5\,V, $I_s$\,=\,100\,pA.}
\end{figure*}

Here we address these issues using STM measurements of bulk \FeBiSe, coupled with density functional theory (DFT). Previous experiments focused on the magnetic properties of bulk Fe dopants via transport \cite{ChaNanoLett2010}, ARPES \cite{ChenScience2010}, or $\mu$SR and magnetization measurements \cite{SalmanArxiv2012}; or adsorbed Fe \cite{WrayNatPhys2010, ScholzPRL2012, SchlenkPRL2013, HonolkaPRL2012}; or Fe dopants in thin films \cite{SongPRB2012}. In this first STM study of bulk \FeBiSe, we report multiple dispersing QPI modes, which give the first access to the full high energy band structure, up to 1\,eV above the Dirac point. Furthermore, we provide the first real space evidence for the coexistence of QWS and TSS on the pristine surface of a bulk material. Finally, we demonstrate the absence of backscattering of the QWS, and we report the relationship of scattering between TSS, QWS, and bulk. All of these observations are enabled by interstitial Fe in the van der Waals (vdW) gap between adjacent \BiSe\ quintuple layers (QLs).

We studied \FeBiSe\ single crystals with nominal $x = 0$ \cite{EtoPRB2010} and 0.07 \cite{SalmanArxiv2012}. Samples were cleaved in vacuum at $T\sim40$\,K and immediately inserted into our home-built STM at 4\,K. Topographic images were obtained in constant current mode ($I_s$) at a fixed sample bias ($V_s$). Differential conductance $dI/dV$, proportional to the local density of states, was measured at a fixed tip-sample distance using a standard lockin technique. Theoretical calculations were performed using the linearized augmented-plane-wave method in the WIEN2K packages \cite{wien2k} and an {\it ab initio} DFT slab method \cite{ZhuPRL2013}.

Topographic images of flat terraces of \FeBiSe\ reveal multiple species of three-fold symmetric defects, segregated into three distinct micron-scale regions, and typified in Figs.\,\ref{fig:impurities}(a-c). Region I is dominated by the same Se vacancies and Bi$_{\mathrm{Se}}$ antisite defects seen in ``pristine'' \BiSe\ \cite{UrazhdinPRB2002, HorPRB2009}. Region II is dominated by Fe substitutions in the top two Bi layers, Fe$_{\mathrm{Bi1}}$ and Fe$_{\mathrm{Bi2}}$ \cite{SongPRB2012}. Region III is dominated by two larger defects, detailed in Figs.\,\ref{fig:impurities}(d-e). The first (``type-A''), centered between topmost Se atoms, was identified as an interstitial Fe \cite{SongPRB2012}. The second (``type-B''), centered on a topmost Se atom, has not been previously observed. Based on their vertex-to-vertex distance of $6a_0 = 2.4$\,nm, and the assumption that their perturbations propagate primarily along the $pp\sigma$ chains extending out and upwards to the surface \cite{UrazhdinPRB2002}, we conclude that both defects in region III are Fe interstitials at inequivalent sites within the vdW gap beneath the top QL. The density of atomic Fe defects in regions II and III is lower than the nominal doping by a factor of $\sim100$, but occasional Fe clusters do appear \cite{Supplement}.

The low Fe concentration is consistent with the small size of Fe$^{3+}$ compared to Bi$^{3+}$, which makes it difficult to dope \cite{ZhangPRL2012}. The spatial separation of distinct impurities can be understood from calculations showing that Fe substitution in \BiSe\ is allowed only in Se-rich conditions \cite{ZhangPRL2012, AbdallaPRB2013}, which also inhibit the formation of Se vacancies \cite{WestPRB2012, WangPRB2013}. On the other hand, Fe interstitials are not energetically favored under any Bi/Se ratio. Such interstitials may appear due to kinetic constraint where the local growth temperature is too low for Fe to overcome the energy barrier to substitute for Bi \cite{ZhangPRL2012}.

\begin{figure*} [t]
 \includegraphics[width=2\columnwidth,clip]{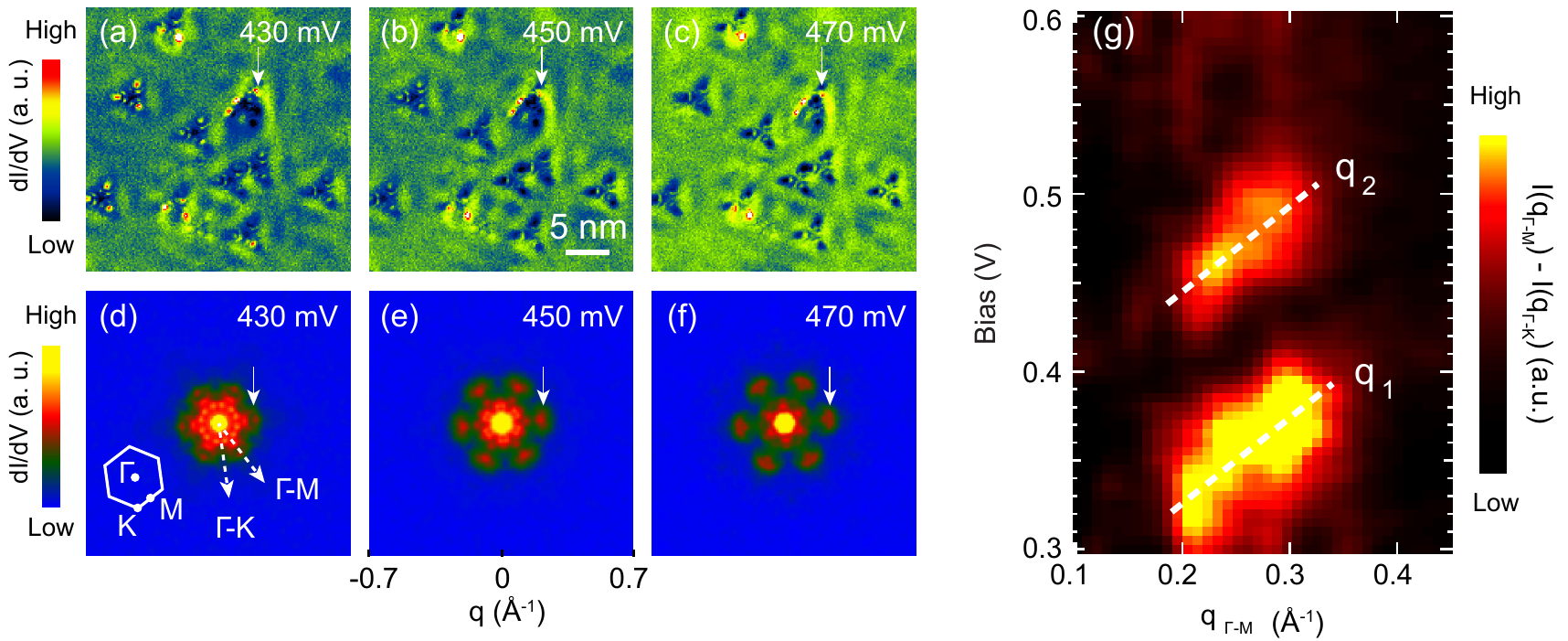}
  \caption {\label{fig:qpi} (color online). (a-c) $dI/dV$ maps over a $30\times30$ nm$^2$ region at the indicated bias. (d-f) Fourier transform (FT) $dI/dV$ maps at varying bias over a $60\times60$ nm$^2$ region encompassing the area of (a)-(c).  A white arrow points to the dispersive \GM\ scattering mode. Inset in (d) is a schematic of the surface Brillouin zone with the high symmetry points. Data was acquired at $T=6$ K, $V_s = 0.6$ V, $I_s = 300$ pA, $V_\text{rms} = 7$ mV (a-c), $V_\text{rms} = 5$ mV (d-f). (g) Fourier linecut showing $q_1$ and $q_2$ scattering modes. The Fourier intensity of the linecut in the \GK\ direction is subtracted from the Fourier intensity of the linecut in the \GM\ direction for the FT-$dI/dV$ maps shown in (d)-(f). The white-dashed lines highlight the linear dispersions of $q_1$ and $q_2$.}
\end{figure*}

\begin{figure} []
 \includegraphics[width=1\columnwidth,clip]{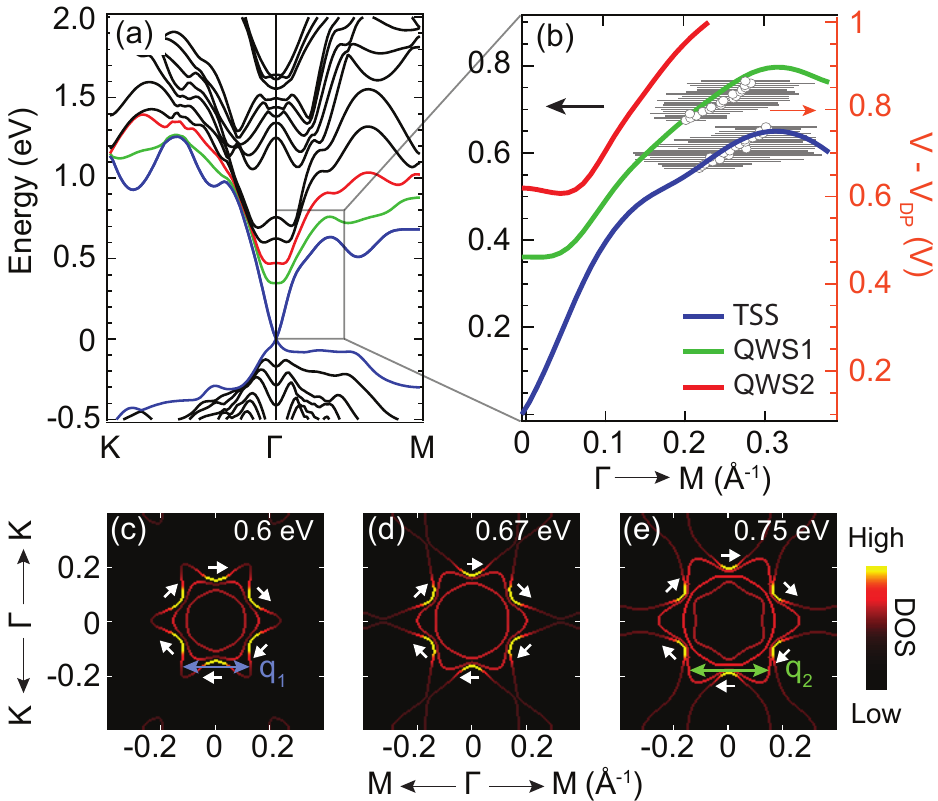}
 \caption {\label{fig:qws} (color online). (a) Calculated band structure of a 5QL slab with the TSS and first two QWS in blue, green and red, respectively. (b) Comparison between the calculated high-energy \GM\ bands and the measured dispersions of the two QPI modes. The grey circles and horizontal bars represent the centers and widths of Gaussian fits to the intensity profile at each bias in Fig.\,\ref{fig:qpi}(g). The measured Dirac point bias $V_{\mathrm{DP}}$ is -0.35\,V \cite{Supplement}. The energy offset between left and right axes represents a rigid band shift which arises from our approximation of the near-surface potential gradient as a square well in our DFT calculations. (c-e) Constant energy contours at energies spanning the transition from dominant TSS scattering ($q_1$) to dominant QWS1 scattering ($q_2$). The color scale is defined by the density of states of the top five atomic layers in a 5QL slab. The expected spin texture due to Rashba-like spin splitting (not included in the calculation) is sketched by white arrows.}
\end{figure}

We now focus on the electronic structure in region III. The $dI/dV$ maps in Figs.\,\ref{fig:qpi}(a-c) display clear QPI patterns, particularly around the new type-B interstitials. The QPI has three remarkable features. First, the scattering modes appear at unexpected high energy levels, almost 1\,eV above the Dirac point. At such high energies the band structure of \BiSe\ was expected to consist of bulk continuum states and was not expected to have a well-defined TSS \cite{ZhangPRL2009, EremeevJETP2010}. Second, the Fourier transform $dI/dV$ maps in Figs.\,\ref{fig:qpi}(d-f) show no dispersing \GK\ modes, but two \GM\ modes ($q_1$ and $q_2$) which disperse with similar slope but $\sim$0.1\,eV relative offset. To better isolate the dispersing \GM\ modes from the long-wavelength dopant disorder (presumed to be isotropic), we plot $I(q_{\Gamma-M}) - I(q_{\Gamma-K})$ as a function of bias voltage $V_s$ and wavevector $q$ in Fig.\,\ref{fig:qpi}(g). Third, the \GM\ mode velocities, $v_{qM} \sim 1.1 - 1.3\,\text{eV\,\AA}$, are substantially smaller than the ARPES-measured TSS velocity near the Dirac point ($v_{\mathrm{DP}} = 3.5\,\text{ eV\,\AA}$) \cite{KurodaPRL2010}. Finally, we emphasize that QPI is observed only in region III.

We argue that the parallel \GM\ scattering modes in Fig.\,\ref{fig:qpi}(g) demonstrate a remarkable extension of both TSS and QWS, far above the energy range accessible to ARPES \cite{BianchiNatComm2010, WrayNatPhys2010, BeniaPRL2011, ZhuPRL2011, BianchiPRL2011, ChenPNAS2012, VallaPRL2012, RoyPRL2014, KingPRL2011, BahramyNatCom2012, YeArxiv2011}. To support this claim, Fig.\,\ref{fig:qws}(a) shows our calculated band structure in a 5QL slab, the minimal system which prevents interactions between top and bottom surfaces, but also supports QWS with depth comparable to fits of ARPES data \cite{BianchiNatComm2010, BeniaPRL2011, KingPRL2011} and the independently measured bulk screening length \cite{LoptienPRB2014}. (We also checked that the calculated band velocity here is a robust feature, nearly independent of the slab thickness \cite{Supplement}.) Constant energy contours (CECs) in Figs.\,\ref{fig:qws}(c-e) show the hexagonal warping \cite{FuPRL2009} which emerges at high energy in both TSS and QWS, along with the expected spin texture of these states \cite{BahramyNatCom2012}. Similar to previous studies on \BiTe\ \cite{ZhangPRL2009, AlpichshevPRL2010, SessiPRB2013}, scattering in \BiSe\ is expected to depend on the warping of the CECs, with a dominant mode emerging between two adjacent $M$-oriented corners of the hexagram (denoted by $q_M$). Because these corners are separated by an angle of $\pi/3$, the velocity of $q_{M}$ should well approximate the band velocity along the \GM\ direction: $v_{q_M} = v_{k(\Gamma-M)}$ \cite{Supplement}. Furthermore, the warping of TSS and QWS in Figs.\,\ref{fig:qws}(c-e) evolves similarly, but shifted in energy, akin to the parallel scattering modes observed in Fig.\,\ref{fig:qpi}(g). Indeed, Fig.\,\ref{fig:qws}(b) shows excellent quantitative agreement between our STM scattering data and DFT calculations, confirming that $q_1$ and $q_2$ modes arise from scattering between adjacent $M$-oriented corners of the CECs of the TSS and QWS1, respectively.

\begin{figure}[]
 \includegraphics[width=1\columnwidth,clip]{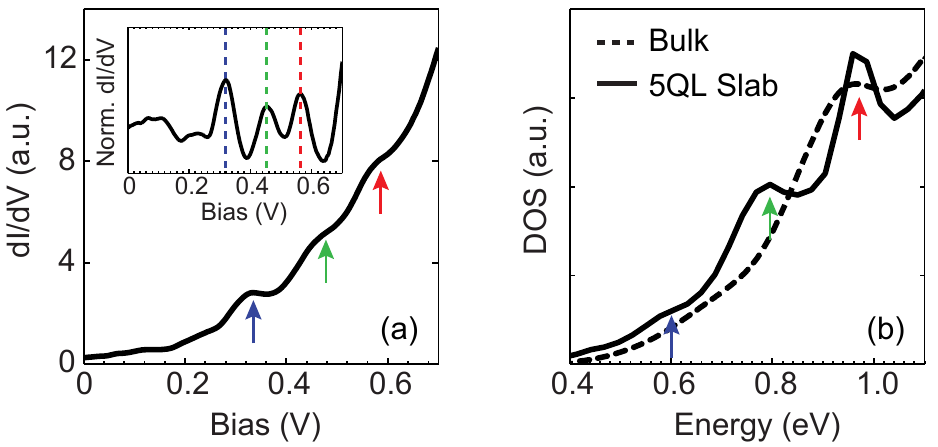}
  \caption{\label{fig:dos} (color online). (a) Spatially averaged $dI/dV$ spectrum ($V_s$\,=\,0.7\,V, $I_s$\,=\,0.34\,nA, and $V_{\mathrm{rms}}$\,=\,5.6\,mV) showing three prominent kinks at 0.32, 0.45 and 0.56\,V. The inset shows the spectrum after subtracting a third order polynomial background. (b) Calculated density of states (DOS) from the bulk (dashed line) and a 5QL slab (solid line) of \BiSe\ (with $V_{\mathrm{DP}}=0$ in the calculation). The slab DOS profile has kinks similar to those observed in the experimental spectrum in (a).}
\end{figure}

Spectroscopic evidence further supports the presence of QWS. In Fig.\,\ref{fig:dos}(a), the spatially averaged $dI/dV$ spectrum from region III shows three distinct kinks over the same energy range as the observed QPI. These kinks are qualitatively well reproduced in Fig.\,\ref{fig:dos}(b), which shows the momentum-integrated density of states from the DFT band structure in Fig.\,\ref{fig:qws}(a). By comparison with Fig.\,\ref{fig:qws}(a), we see that the three kinks in the DOS occur near the three energies where the linear dispersions of the TSS, QWS1, and QWS2 bands break down.

Despite the breakdown of linear dispersion and the onset of bulk bands, both TSS and QWS remain unexpectedly robust against backscattering and surface-bulk scattering. We observe no dispersing \GK\ modes throughout the measured energy range, which rules out backscattering and indicates the spin polarization of the bands to almost 1\,eV above the Dirac point. We observe one weak, non-dispersing \GK\ mode around $V_s=300-400$\,meV above \cite{Supplement}. Previous studies ascribed non-dispersing QPI modes to surface-bulk scattering in \BiSe\ \cite{KimPRL2011}. However, our observed wavevector $q_K\sim 0.44\,\mathrm{\AA}^{-1}$ corresponds very nearly to the $2\sqrt{3}a$ distance from the bright center to the lobes of the type-B interstitial Fe defects, which is determined simply by the bond geometry of the impurity \cite{UrazhdinPRB2002}.
We therefore find no evidence of backscattering or surface-bulk scattering deteriorating the TSS or the QWS at high energy in \BiSe.

Finally, we comment on the origin of QWS. Previously observed QWS in \BiSe\ were believed to arise from band bending due to charged adsorbents \cite{BeniaPRL2011, ZhuPRL2011, BianchiPRL2011, KingPRL2011, BahramyNatCom2012}, or vdW gap expansion due to burrowing adsorbents \cite{YeArxiv2011, EremeevNJP2012}. Here we have reported the first evidence of QWS at the clean surface of a bulk crystal. Fe interstitials may induce QWS at the clean surface of bulk \FeBiSe\ by two mechanisms: (1) their ionization to Fe$^{3+}$ would donate three electrons to induce downward band bending \cite{SongPRB2012}; (2) their cores would physically expand the vdW gap. Either mechanism supposes vertical dopant inhomogeneity, in keeping with the lateral inhomogeneity demonstrated in Fig.\,\ref{fig:impurities}. (The tip itself can't be responsible for the the observed QWS, because the tip's positive bias can't cause the downward band bending required to induce QWS in the bulk conduction band.)

In conclusion, our STM/STS study of \FeBiSe\ provides fundamental new information about the empty state band structure of \BiSe. Scattering from a previously unobserved interstitial Fe defect allowed QPI imaging and band structure reconstruction up to 1\,eV above the Dirac point, from which we draw three main conclusions. First, quantitative comparison between our QPI data and DFT calculations proves the unexpected appearance of QWS, coexisting with TSS, at the clean surface of a bulk material. These well-defined surface states persist to almost 1\,eV above the Dirac point, overlapping by $\sim0.6$\,eV with the bulk conduction band, but showing no evidence of backscattering. Second, distinctive kinks in $dI/dV$ spectra reveal the breakdown of linear dispersion for TSS and the first two QWS. The new high-energy velocity is $3\times$ smaller than the velocity near the Dirac point, and is consistent with our DFT calculations. Third, the absence of dispersing \GK\ scattering demonstrates the protection of both TSS and QWS against backscattering, well beyond the breakdown of linear dispersion, even in the presence of sparse Fe dopants. Our results suggest that interstitial defects in the vdW gap of \BiSe\ can lead to the coexistence of protected spin-polarized QWS and TSS over a wide energy range, far outside the bulk band gap. These discoveries bode well for the controlled use of spins in future devices.

\acknowledgements
We benefited from computational resources provided by Ilya Elfimov. This research was supported by the US National Sciences Foundation under grant DMR-110623. M.M.Y. was supported by an NSERC PGS-D fellowship, Z.H.Z. was supported by an NSERC PDF fellowship, A.S. was supported by A*STAR, Y.H. was supported by the New York Community Trust - George Merck fund, and C.L.S. was supported by the Lawrence Golub fellowship at Harvard University. Work at Osaka University was supported by JSPS (KAKENHI 25220708) and AFOSR (AOARD 124038).

\bibliography{FeBiSe}




\widetext
\clearpage

\linespread{1.2}






\begin{center}

{\LARGE Supplementary Material for:}
\vspace*{0.35cm}
\par
{\Large\textbf{Spin-Polarized Quantum Well States on Bi$_{2-x}$Fe$_x$Se$_3$}}
\vspace*{0.35cm}
\par
\vspace*{0.35cm}
\par
\large{Michael\,M.\,Yee,
Z.\,-H.\,Zhu,
Anjan\,Soumyanarayanan,
Yang\,He,
Can-Li\,Song,
Ekaterina\,Pomjakushina,
Zaher\,Salman,
Amit\,Kanigel,
Kouji\,Segawa,
Yoichi\,Ando,
Jennifer\,E.\,Hoffman}

\end{center}

\setcounter{equation}{0}
\setcounter{figure}{0}
\setcounter{table}{0}
\setcounter{page}{1}
\renewcommand{\thefigure}{S\arabic{figure}}

\vspace{1.5cm}

\begin{figure}[h!]
\centering
 \includegraphics[width=0.6\columnwidth]{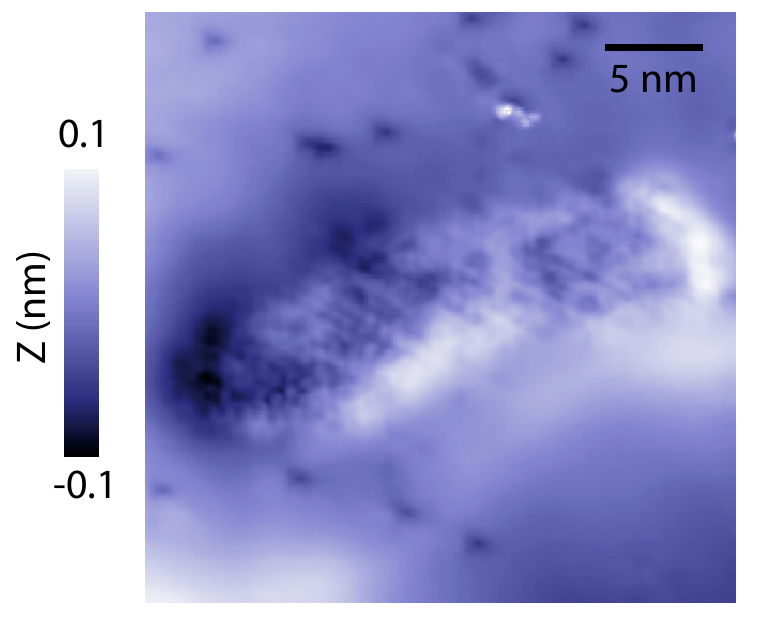}
  \caption[Fe clustering in \FeBiSe\ ]
  {An example of occasionally observed Fe clustering in \FeBiSe\ crystals. The 30 $\times$ 30 nm$^2$ topography was acquired at $T=7$\,K, $V_s = 0.3$\,V, $I_s = 30$\,pA.
\label{fig:cluster}
}
\end{figure}

\begin{figure}[h!]
 \includegraphics[width=\columnwidth,clip]{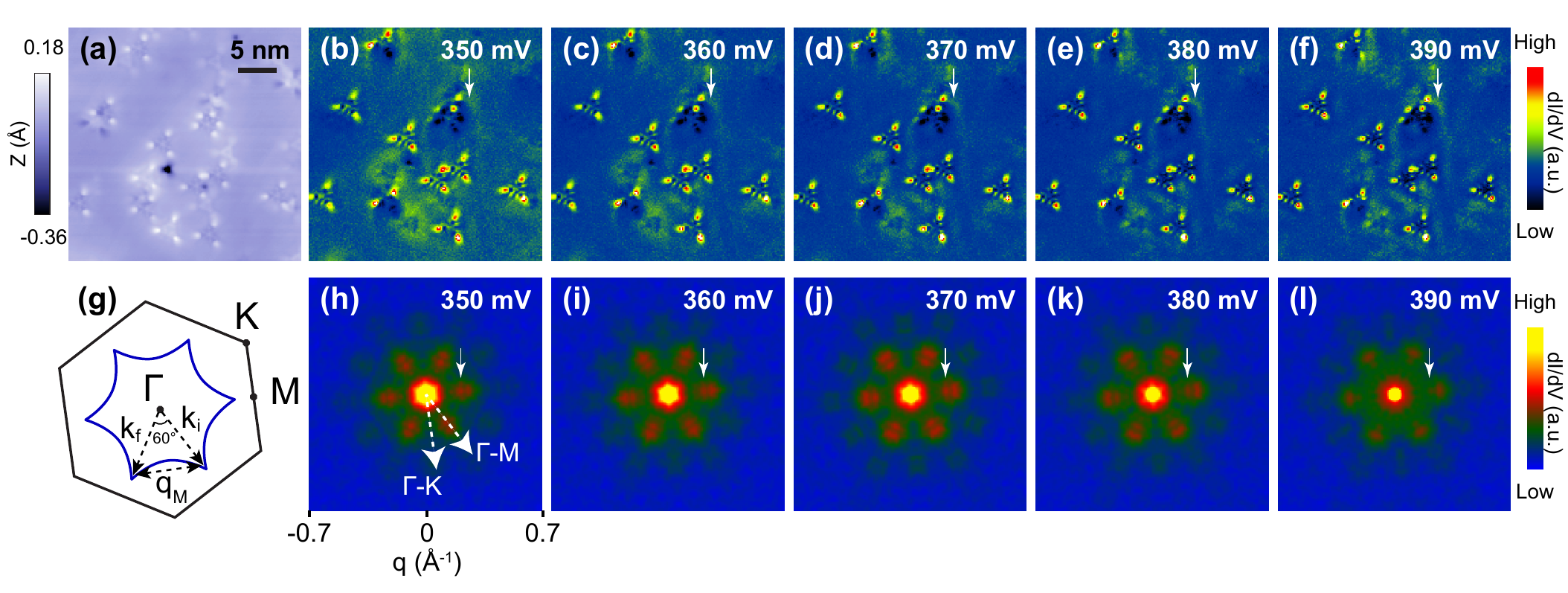}
  \caption{ (a) Topography of a $30\times30$ nm$^2$ region of \FeBiSe\ ($T=6$\,K, $V_s=0.6$\,V, $I_s=30$\,pA). (b-f) $dI/dV$ maps over the region shown in (a) at the indicated bias. Interference waves are seen around the type-B defects. A white arrow is overlaid on the five images and shows that as the energy is increased the wavefront moves towards the impurity ($T=6$\,K, $V_s=0.6$\,V, $I_s=300$\,pA, $V_{\mathrm{rms}}=7$\,mV). (g) Schematic diagram of the Brillouin zone (black) and a hexagonally warped constant energy contour appropriate for \BiSe. The magnitude of scattering vector $q_M$ should be similar to the value of initial and final quasiparticle wavevectors $k_i$ and $k_f$ because of the $\pi$/3 angle separation. Thus, the velocity of $q_M$ is expected to be equal to the \GM\ band velocity. (h-l) Fourier transforms of the $dI/dV$ maps at varying bias over $60\times60$ nm$^2$ regions encompassing the area of (a-f) ($T=6$\,K, $V_s=0.6$\,V, $I_s=300$\,pA, $V_{\mathrm{rms}}=5$\,mV). A white arrow is overlaid to show that as energy increases the scattering wavevector increases in the \GM\ direction.
\label{fig:qpi_q1}
}
\end{figure}

\begin{figure}[h!]
\centering
 \includegraphics[width=0.75\columnwidth,clip]{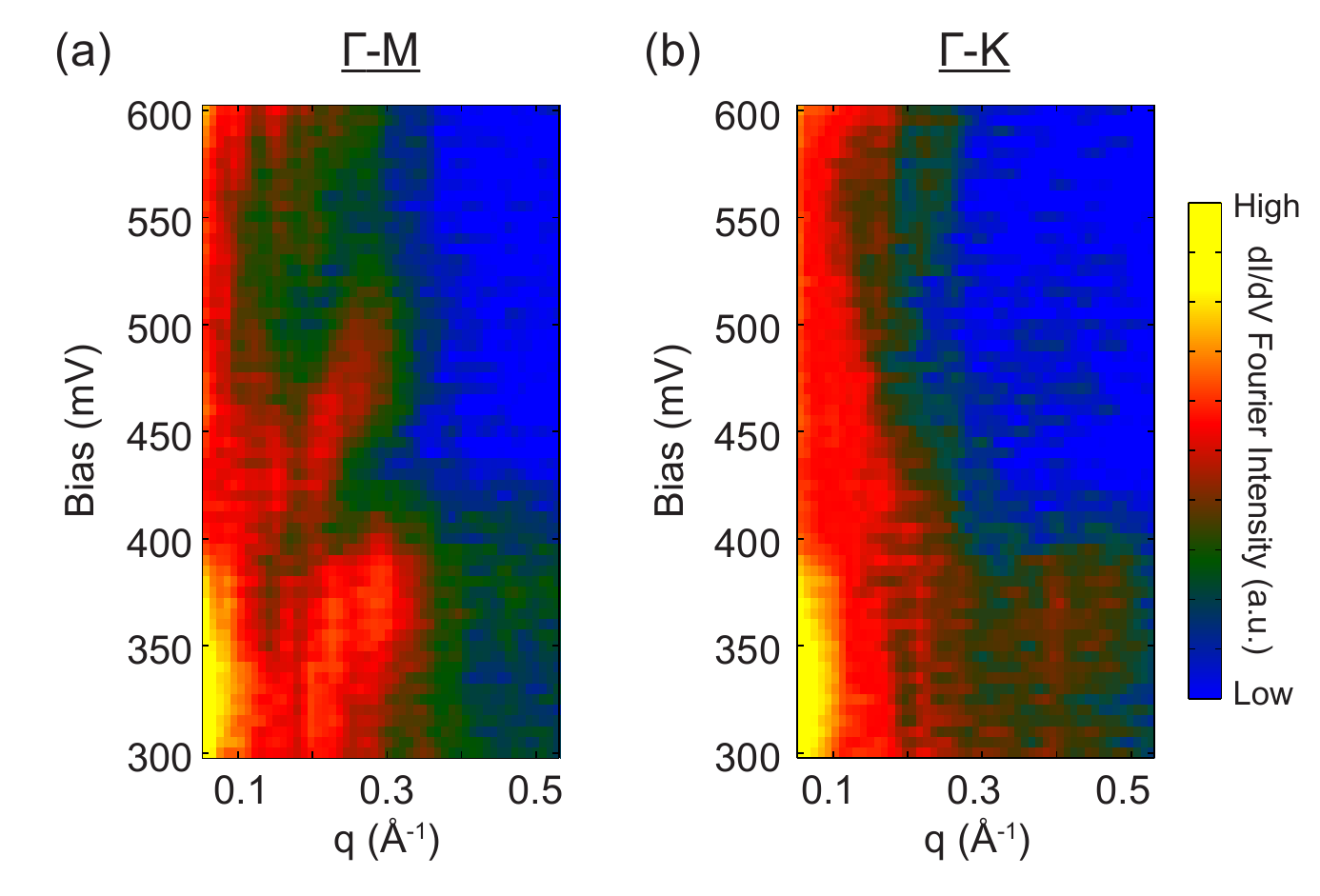}
  \caption{ (a, b) Linecuts of Fourier transform $dI/dV$ spectroscopy maps shown in Fig.\,2 of the main text and Fig.\,\ref{fig:qpi_q1}. Linecuts are taken in the (a) \GM\ and (b) \GK\ directions. The difference of the two images gives the result shown in Fig.\,2(g) of the main text, which highlights the $q_1$ and $q_2$ \GM\ scattering modes. The $dI/dV$ maps shown in Fig.\,\ref{fig:qpi_q1} and Fig.\,2 of the main text were normalized by energy layer, Fourier transformed, six-fold symmetrized, and mirrored along a Bragg direction.
\label{fig:GMGKlinecuts}
}
\end{figure}

\begin{figure}[h!]
\centering
\includegraphics[width = 0.5\columnwidth] {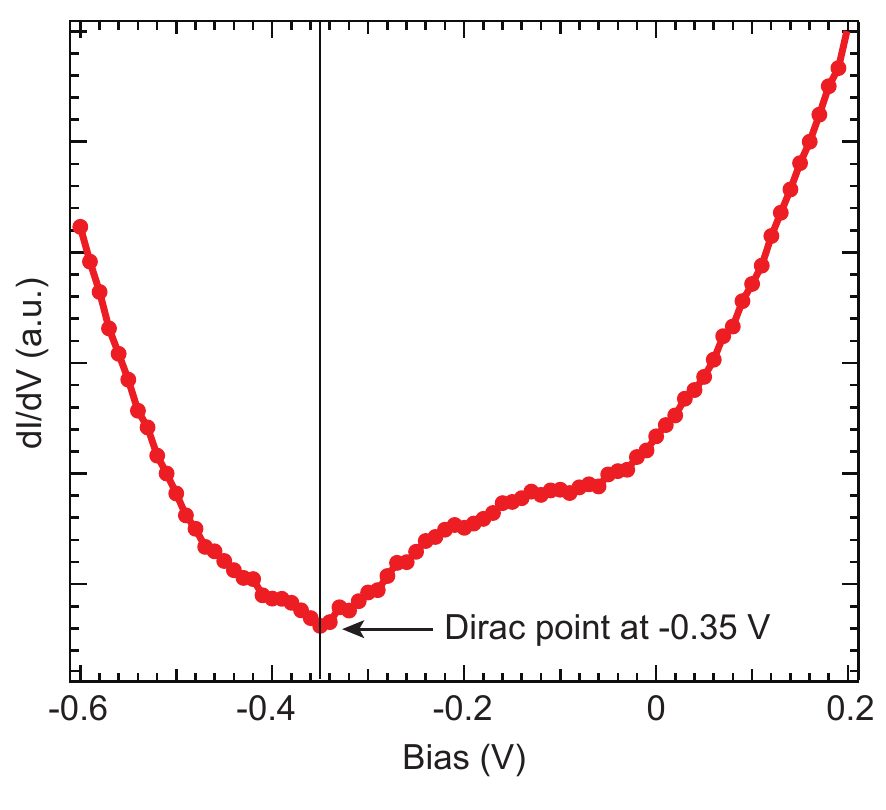}
\caption {
$dI/dV$ spectrum acquired in region III with $V_s$\,=\,0.5\,V and $I_s$\,=\,100\,pA shows a Dirac point at $-0.35$\,V.
\label{fig:dp}
}
\end{figure}

\begin{figure}[h!]
\centering
\includegraphics[width = 0.65\columnwidth] {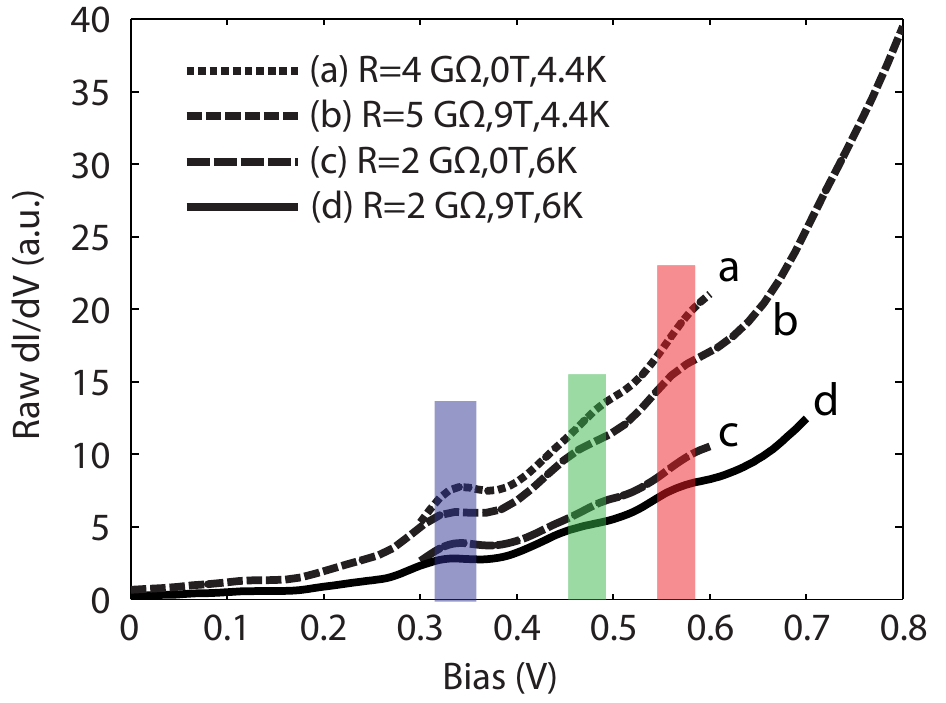}
\caption {
$dI/dV$ spectra were acquired at the listed experimental conditions, with varying junction resistance $R = V_s/I_s$. The junction resistance is exponentially dependent on the tip-sample separation. Blue, green, and red vertical bars are overlaid to show the approximate energies of kinks in the spectra, which we identify as the breakdowns of linear dispersion of the TSS, QWS1, and QWS2, respectively. The kink energies are consistent across spectra taken with different experimental conditions, both in and out of magnetic field. Experimental conditions not listed in the legend: (a) $V_s = 400$ mV, $V_{\mathrm{rms}} = 14$ mV; (b) $V_s = 500$ mV, $V_{\mathrm{rms}} = 14$ mV;  (c) $V_s = 600$ mV, $V_{\mathrm{rms}} = 5$ mV; (d) $V_s = 700$ mV, $V_{\mathrm{rms}} = 5.6$ mV. Spectrum (d) is the same data presented in Fig.\,4 of the main text.
\label{fig:bb_exp}
}
\end{figure}

\begin{figure}[h!]
\centering
 \includegraphics[width=0.95\columnwidth,clip]{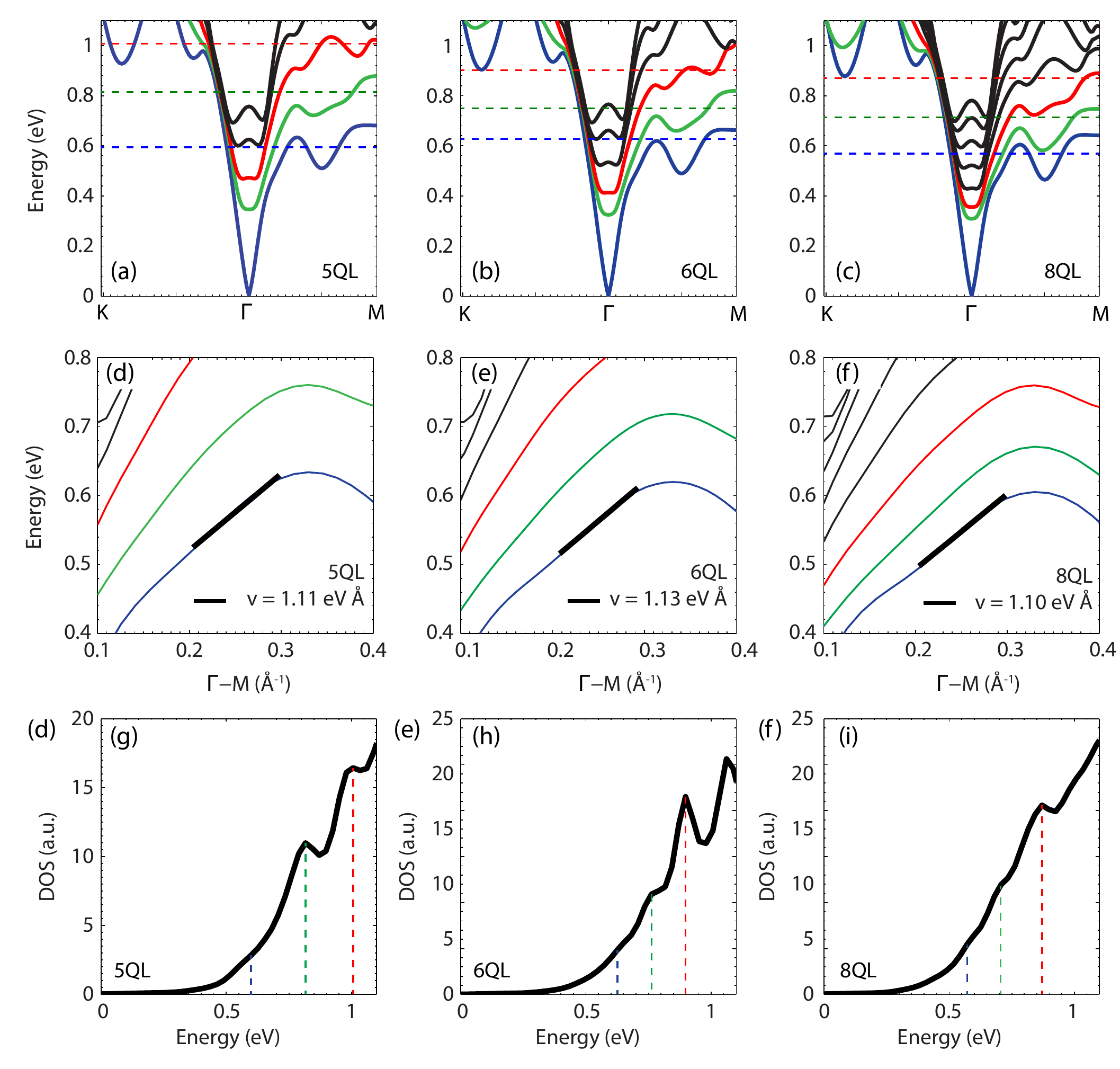}
  \caption {(a-c) DFT slab calculations with various QL thicknesses. (d-f) Band structure along the \GM\ direction at the indicated QL slab thickness, in an energy and $k$-space range corresponding to the $q_1$ and $q_2$ scattering modes in the data. The black line shows the best linear fit ($E = vk + E_0$) in the range $k = 0.2$ \AA$^{-1}$ to $k=0.3$ \AA$^{-1}$. The velocity extracted from the fit is equal to $\sim 1.1\,\text{eV\AA}$ and is nearly independent of the slab thickness. (g-i) Calculated DOS at varying QL thicknesses. Dashed vertical lines in (g)-(i) are guides illustrating the energies of the kinks in the DOS. These kink energies are also marked as horizontal lines in (a)-(c). Although a full Poisson-Schr\"{o}dinger calculation of the potential gradient profile \cite{BianchiNatComm2010, KingPRL2011} is beyond the scope of this work, the calculated DOS kinks arising from the simple square well potential with thickness 5QL, 6QL, or 8QL all show excellent qualitative agreement with the measured data in main text Fig.\ 4a.
\label{fig:dft_varyN}
}
\end{figure}

\end{document}